\documentclass[seceq]{ptptex}

\usepackage{graphicx}

\newcommand{\be}{\begin{equation}}
\newcommand{\ee}{\end{equation}}
\newcommand{\bea}{\begin{eqnarray}}
\newcommand{\eea}{\end{eqnarray}}
\newcommand{\lm}{\Lambda}
\newcommand{\vlowk}{V_{{\rm low}\,k}}

\newcommand{\fmi}{\, \text{fm}^{-1}}

\newcommand{\mev}{\, \text{MeV}}

\newcommand{\kf}{k_{\rm F}}

\newcommand{\openone}{\leavevmode\hbox{\small1\normalsize\kern-.33em1}}

\title{
Neutron Matter from Low-Momentum Interactions%
}

\author{
Bengt \textsc{Friman}$^{1,}$\footnote{E-mail address:
b.friman@gsi.de},
Kai \textsc{Hebeler}$^{1,}$\footnote{E-mail address:
k.hebeler@gsi.de},
Achim \textsc{Schwenk}$^{2,}$\footnote{E-mail address:
schwenk@triumf.ca}
and
Laura \textsc{Tol\'os}$^{3,}$\footnote{E-mail address:
tolos@fias.uni-frankfurt.de}
}

\inst{
$^1$GSI, Planckstr.~1, D-64291 Darmstadt, Germany\\
$^{2}$TRIUMF, 4004 Wesbrook Mall, Vancouver, BC, Canada, V6T 2A3\\
$^{2}$ FIAS, J.W.~Goethe Universit\"at, D-60438 Frankfurt am Main, Germany}

\abst{We present a perturbative calculation of the neutron matter
equation of state based on low-momentum two- and three-nucleon
interactions. Our results are compared to the model-independent
virial equation of state and to variational calculations, and we
provide theoretical error estimates by varying the cutoff used to
regulate nuclear interactions. In addition, we study the dependence
of the BCS $^1$S$_0$ superfluid pairing gap on nuclear interactions
and on the cutoff. The resulting gaps are well constrained by the
nucleon-nucleon scattering phase shifts, and the cutoff dependence
is very weak for sharp or sufficiently narrow smooth regulators
with cutoffs $\lm > 1.6 \fmi$. }

\begin{document}

\maketitle

\section{Introduction}

The determination of a reliable equation of state of nucleonic
matter plays a central role for the physics of neutron
stars~\cite{LP} and core-collapse supernovae~\cite{Mezza,Janka}.
Furthermore the superfluidity and superconductivity of neutrons and
protons is an important phenomenon in nuclear many-body
systems~\cite{Litvinov,Sarazin}, in particular for the cooling of
neutron stars~\cite{Yakovlev}. In this contribution, we present
calculations of the neutron matter equation of state at finite
temperature and of the $^1$S$_0$ superfluid gap in the BCS
approximation based on low-momentum interactions.

Renormalization group methods coupled with effective field theory
(EFT) offer the possibility for a systematic approach to the
equation of state. By evolving nuclear forces to low-momentum
interactions $V_{{\rm low}\,
k}$~\cite{VlowkReport,Vlowknucmatt,VlowkWeinberg} with cutoffs
around $2 \, {\rm fm}^{-1}$, the model-dependent short-range
repulsion is integrated out and the resulting low-momentum
interactions are well constrained by the nucleon-nucleon (NN)
scattering data. Furthermore, the corresponding leading-order
three-nucleon (3N) interactions (based on chiral EFT) become
perturbative in light nuclei for $\lm \lesssim 2 \, {\rm
fm}^{-1}$~\cite{Vlowk3N}.

With increasing density, Pauli blocking eliminates the shallow
two-nucleon bound and nearly-bound states, and the contribution
of the particle-particle channel to bulk properties becomes
perturbative in nuclear matter~\cite{Vlowknucmatt}. The Hartree-Fock
approximation is then a good starting point for many-body
calculations with low-momentum NN and 3N interactions, and
perturbation theory (in the sense of a loop expansion) around the
Hartree-Fock energy converges at moderate densities. This can be
understood quantitatively based on the behavior of the Weinberg
eigenvalues as a function of the cutoff and
density~\cite{VlowkWeinberg,Vlowknucmatt}.

Some uncertainty remained concerning a possible dependence of the
$^1$S$_0$ pairing gap on the input NN interaction in low-density
neutron matter ($\kf < 1.6 \fmi$). We address this point and
explore the dependence of $^1$S$_0$ superfluidity on nuclear
interactions at the BCS level in detail. We find that the BCS gap
is well constrained by the NN phase shifts. Therefore, any
uncertainties are due to polarization (induced interaction),
dispersion and three-nucleon interaction effects.

\section{Equation of State of Neutron Matter}

Using the Kohn-Luttinger-Ward theorem~\cite{KL,LW}, the
perturbative expansion of the free energy (at finite temperature)
can be formulated as a loop expansion around the Hartree-Fock (HF)
energy. In this work, we include the first-order NN and 3N
contributions, as well as normal and anomalous second-order NN
diagrams. Other thermodynamic quantities are computed using
standard thermodynamic relations.

\begin{figure}
\begin{center}
\includegraphics[clip=,width=12cm]{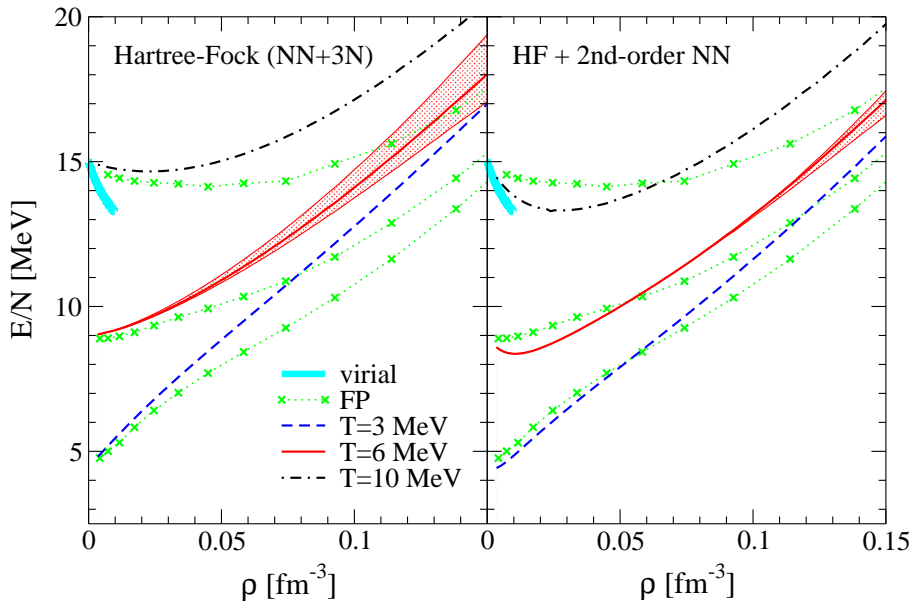}
\end{center}
\caption{Energy per particle $E/N$ as a function of the density $\rho$
at first order (left panel) and including second-order NN contributions
(right panel).~\cite{finiteT}}
\label{en}
\end{figure}

The resulting energy per particle $E/N$ as a function of the
density $\rho$ is shown in Fig.~\ref{en} for a cutoff $\lm = 2.1
\fmi$ and temperatures $T=3$, $6$ and $10 \mev$~\cite{finiteT}. The
results presented in the left panel are the first-order NN and 3N
contributions, and those in the right panel includes all
second-order diagrams with NN interactions. For $T=6 \mev$, we also
give a band spanned between $\Lambda=1.9 \fmi$ (lower line) and
$\Lambda= 2.5 \fmi$ (upper line). The inclusion of second-order
contributions significantly reduces the cutoff dependence
of the results. The model-independent virial equation
of state~\cite{vEOSneut} and the variational calculations of
Friedman and Pandharipande (FP)~\cite{FP} are displayed for
comparison.

The inclusion of second-order correlations lowers the energy below
the variational results for densities $\rho \lesssim 0.05 \, {\rm
fm}^{-3}$, and we observe a good agreement for $E/N$ with the $T=10
\mev$ virial result when the second-order contributions are
included. In the virial equation of state these contributions are
included via the second-order virial coefficient, while in the
variational calculation the state dependence of such correlations
is only partly accounted for.~\cite{FFP} Furthermore, the generic
enhancement of the effective mass at the Fermi surface leads to an
enhancement of the entropy at low temperatures above the
variational and HF results.~\cite{finiteT,FFP,FPS}

\section{BCS gap in the $^1$S$_0$ channel}

We solve the BCS gap equation in the $^1$S$_0$ channel
\begin{equation}
\Delta(k) = - \frac{1}{\pi} \int dp \, p^2 \: \frac{V_{{\rm low}\,k}(k,p) \,
\Delta(p)}{\sqrt{\xi^2(p) + \Delta^2(p)}} \,,
\label{gapeq}
\end{equation}
with the (free-space) low-momentum NN interaction $V_{{\rm
low}\,k}(k,k')$. Here $\xi(p) \equiv
\varepsilon(p) - \mu$, $\varepsilon(p)=p^2/2$ and
$\mu = \kf^2/2$ ($c=\hbar=m=1$).

We find that the neutron-neutron BCS gap is practically independent
of the NN interaction~\cite{BCS}. Consequently, $^1$S$_0$
superfluidity is strongly constrained by the NN scattering phase
shifts. The maximal gap at the BCS level is $\Delta \approx 2.9-3.0
\mev$ for $\kf\approx 0.8-0.9 \fmi$. For the neutron-proton
$^1$S$_0$ case, we find somewhat larger gaps, reflecting the charge
dependence of realistic nuclear interactions.~\cite{BCS}

\begin{figure}[t]
\begin{center}
\includegraphics[clip=,width=8cm]{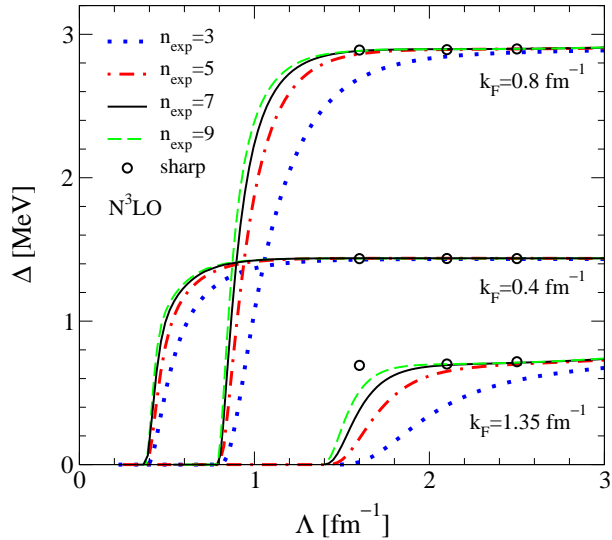}
\end{center}
\caption{The neutron-neutron $^1$S$_0$ superfluid
pairing gap $\Delta$ as a function of the cutoff $\lm$ for three
densities and different smooth exponential regulators, as well as
for a sharp cutoff~\cite{BCS}. The low-momentum interactions are
derived from the N$^3$LO chiral potential of Ref.~\cite{N3LO}.}
\label{flow_nexp}
\end{figure}

In Fig.~\ref{flow_nexp} we show the dependence of the
neutron-neutron $^1$S$_0$ superfluid pairing gap on the cutoff
starting from the N$^3$LO chiral potential of Ref.~\cite{N3LO} for
three representative densities.~\cite{BCS} We employed different
smooth exponential regulators $f(k)=\exp[-(k^2/\lm^2)^n]$, as well
as a sharp cutoff. As long as the cutoff is large compared to the
dominant momentum components of the bound state ($\lm > 1.2 \kf$),
the gap depends very weakly on the cutoff. This shows that the
$^1$S$_0$ superfluid pairing gap probes low-momentum physics. Below
this scale, which depends on the density and the smoothness of the
regulator, the gap decreases, since the relevant momentum
components of the Cooper pair are then partly integrated out.

\section{Conclusions}

In summary, we have studied the equation of state at finite
temperature including many-body contributions in a systematic
approach. We have found good agreement with the virial equation of
state in the low-density--high-temperature regime. Analyzing the
cutoff dependence of our results provides lower bounds for the
theoretical uncertainties. The possibility of estimating
theoretical errors plays an important role for reliable
extrapolations to the extreme conditions reached in astrophysics.

In addition, we have shown that the $^1$S$_0$ superfluid pairing
gap in the BCS approximation is practically independent of the
choice of NN interaction, and therefore well constrained by the NN
scattering data. This includes a very weak cutoff dependence with
low-momentum interactions $\vlowk$ for sharp or sufficiently narrow
smooth regulators with $\lm > 1.6 \fmi$. At lower densities, it is
possible to lower the cutoff further to $\lm > 1.2 \kf$.
Furthermore, the pairing gap clearly reflects the charge dependence
of nuclear interactions. The weak cutoff dependence indicates that,
in the $^1$S$_0$ channel, the contribution of 3N interactions is
small at the BCS level.

\section*{Acknowledgements}

This work was supported in part by the Virtual Institute VH-VI-041
of the Helmholtz Association, NSERC and US DOE Grant
DE--FG02--97ER41014. TRIUMF receives federal funding via a
contribution agreement through NRC.

%



\begin{thebibliography}{99}

\bibitem{LP} J.M.~Lattimer and M.~Prakash, \AJ{550,2001,426}.

\bibitem{Mezza} A.~Mezzacappa, Annu. Rev. Nucl. Part. Sci. \textbf{55}
(2005), 467.

\bibitem{Janka} H.T.~Janka, R.~Buras, F.S.~Kitaura Joyanes, A.~Marek
and M.~Rampp, astro-ph/0405289.

\bibitem{Litvinov} Yu.A.~Litvinov {\it et al.}, \PRL{95,2005,042501}.

\bibitem{Sarazin} F.~Sarazin {\it et al.}, \PRC{70,2004,031302(R)}.

\bibitem{Yakovlev} D.G.~Yakovlev and C.J.~Pethick, Ann. Rev. Astron.
Astrophys. \textbf{42} (2004), 169.

\bibitem{VlowkReport} S.K.~Bogner, T.T.S.~Kuo and A.~Schwenk,
\PRP{386,2003,1}.

\bibitem{Vlowknucmatt} S.K.~Bogner, A.~Schwenk, R.J.~Furnstahl and
A.~Nogga, \NPA{763,2005,59}.

\bibitem{VlowkWeinberg} S.K.~Bogner, R.J.~Furnstahl, S.~Ramanan and
A.~Schwenk, \NPA{773,2006,203}.

\bibitem{Vlowk3N} A.~Nogga, S.K.~Bogner and A.~Schwenk,
\PRC{70,2004,061002(R)}.

\bibitem{KL} W.~Kohn and J.M.~Luttinger, \PR{118,1960,41}.

\bibitem{LW} J.M.~Luttinger and J.C.~Ward, \PR{118,1960,1417}.

\bibitem{finiteT} L.~Tol\'os, B.~Friman and A.~Schwenk, nucl-th/0611070;
and to be published.

\bibitem{vEOSneut} C.J.~Horowitz and A.~Schwenk, \PLB{638,2006,153}.

\bibitem{FP} B.~Friedman and V.R.~Pandharipande, \NPA{361,1981,502}.

\bibitem{FFP} S.~Fantoni, B.L.~Friman and V.R.~Pandharipande, \NPA{399,1983,51}.

\bibitem{FPS} S.~Fantoni, V.R.~Pandharipande and K.E.~Schmidt, \PRL{48,1982,878}.

\bibitem{BCS} K.~Hebeler, A.~Schwenk and B.~Friman, nucl-th/0611024,
Phys. Lett. B (in press).

\bibitem{N3LO} D.R.~Entem and R.~Machleidt, \PRC{68,2003,041001(R)}.

\end{thebibliography}
\end{document}